# A simulation study on image reconstruction in magnetic particle imaging with field-free-line encoding

Kenya Murase

*Department of Medical Physics and Engineering, Division of Medical Technology and Science, Faculty of Health Science, Graduate School of Medicine, Osaka University*

*1-7 Yamadaoka, Suita, Osaka 565-0871, Japan*

Address correspondence to:

Kenya Murase, Dr. Med. Sci., Dr. Eng.
Department of Medical Physics and Engineering, Division of Medical Technology and Science, Faculty of Health Science, Graduate School of Medicine, Osaka University
1-7 Yamadaoka, Suita, Osaka 565-0871, Japan
Tel & Fax: (81)-6-6879-2571,
E-mail: murase@sahs.med.osaka-u.ac.jp

**Abstract**

The purpose of this study was to present image reconstruction methods for magnetic particle imaging (MPI) with a field-free-line (FFL) encoding scheme and to propose the use of the maximum likelihood-expectation maximization (ML-EM) algorithm for improving the image quality of MPI.

The feasibility of these methods was investigated by computer simulations, in which the projection data were generated by summing up the Fourier harmonics obtained from the MPI signals based on the Langevin function. Images were reconstructed from the generated projection data using the filtered backprojection (FBP) method and the ML-EM algorithm. The effects of the gradient of selection magnetic field (SMF), the strength of drive magnetic field (DMF), the diameter of magnetic nanoparticles (MNPs), and the number of projection data on the image quality of the reconstructed images were investigated. The spatial resolution of the reconstructed images became better with increasing gradient of SMF and with increasing diameter of MNPs up to approximately 30 nm. For the diameter greater than approximately 30 nm, the spatial resolution was almost constant. It also became better with decreasing strength of DMF. The image quality was improved by use of the ML-EM algorithm compared to the FBP method, especially when the strength of DMF was weak and the number of projection data was small.

In conclusion, we presented image reconstruction methods for MPI with an FFL encoding scheme and our preliminary results suggest that the ML-EM algorithm will be useful for improving the image quality of MPI.

## 1. Introduction

In 2005, a new imaging method called magnetic particle imaging (MPI) was introduced [1] that allows for imaging the spatial distribution of magnetic nanoparticles (MNPs) such as superparamagnetic iron oxide (SPIO) with high sensitivity, high spatial resolution, and high imaging speed.

MPI uses the nonlinear response of MNPs for detecting their presence in an oscillating magnetic field (drive magnetic field). Spatial encoding is realized by saturating the MNPs almost everywhere except in the vicinity of a special point called the field-free point (FFP) using a static magnetic field (selection magnetic field) [1]. Recently, it has been shown that the sensitivity of MPI could be significantly improved by using a simultaneous encoding scheme [2]. This can be accomplished by scanning the region of interest with a field-free line (FFL) instead of the FFP. More recently, Knopp et al. [3] presented the FFL coil assembly for MPI, which is feasible to manufacture requiring resistive coils roughly the same electrical power as that of an FFP scanner of equal size and gradient performance. In this scheme, the FFL moves rapidly back and forth while rotating slowly. This is similar to the image encoding scheme used in x-ray computed tomography (CT) [4], which motivated us to investigate whether the image reconstruction method used in x-ray CT can be applied to MPI with an FFL encoding scheme.

The filtered backprojection (FBP) method has been mainly used for image reconstruction in x-ray CT, whereas an iterative statistical method such as the maximum likelihood-expectation maximization (ML-EM) algorithm [5] or an accelerated version of the ML-EM algorithm called ordered subsets-expectation maximization (OS-EM) algorithm [6] has been used for image reconstruction of positron emission tomography (PET) or single photon emission computed tomography (SPECT). It has been shown that the ML-EM or OS-EM algorithm is more useful than the FBP method for improving the image quality of PET [7] or SPECT [8].

The purpose of this study was to present image reconstruction methods for MPI with an FFL encoding scheme, to investigate the feasibility of these methods using computer simulations, and to propose the use of the ML-EM algorithm for improving the image quality of MPI.

## 2. Materials and methods

*2.1. Theory*

Assuming a single receive coil with sensitivity [$\sigma_r(\mathbf{r})$] at spatial position $\mathbf{r}$, the changing magnetization induces a voltage according to Faraday's law, which is given by

$$v(t) = -\mu_0 \frac{d}{dt} \int_\Omega \sigma_r(\mathbf{r}) \cdot M(\mathbf{r},t) d\mathbf{r}, \tag{1}$$

where $\Omega$ denotes the volume containing MNPs, $M(\mathbf{r}, t)$ is the magnetization at position $\mathbf{r}$ and time $t$, and $\mu_0$ is the magnetic permeability of vacuum. The receive coil sensitivity [$\sigma_r(\mathbf{r})$] derives from the magnetic field that the coil would produce if driven with a unit current [9].

In the following, the receive coil sensitivity is assumed to be constant and uniform over the volume of interest and to be denoted by $\sigma_0$. Thus, $v(t)$ given by Eq. (1) is reduced to

$$v(t) = -\mu_0 \sigma_0 \frac{d}{dt} \int_\Omega M(\mathbf{r},t) d\mathbf{r}. \tag{2}$$

Neglecting constant factors, we introduce the notation $s(\mathbf{r},t)$ for the MPI signal generated by a point-like distribution of MNPs at position **r**. If the MNP distribution is approximated by a δ distribution, the volume integral vanishes and the magnetization [$M(\mathbf{r},t)$] is determined by the local magnetic field. The MPI signal can then be given by

$$s(\mathbf{r},t) = -\frac{d}{dt} M(\mathbf{r},t). \tag{3}$$

To consider the noise in MPI signal, Gaussian noise was added to $s(\mathbf{r},t)$ to give a certain signal-to-noise ratio (SNR) in this study. The SNR was given by $SNR = s(\mathbf{r},t)/\sigma$, where $\sigma$ is the standard deviation of the noise generated from normally distributed random numbers with zero mean and unit variance.

*2.2. Langevin function*

The magnetization of MNPs in response to an applied magnetic field can be described by the Langevin function [10], which is given by

$$M(\xi) = M_0(\coth \xi - \frac{1}{\xi}), \quad (4)$$

where $M_0$ is the saturation magnetization and $\xi$ is the ratio between magnetic energy of a particle with magnetic moment $m$ in an external magnetic field $H$, and thermal energy given by the Boltzmann constant $k_B$ and absolute temperature $T$:

$$\xi = \mu_0 m H / k_B T = \mu_0 M_d V_M H / k_B T. \quad (5)$$

In Eq. (5), $M_d$ is the domain magnetization of a suspended particle and $V_M$ is the magnetic volume given by $V_M = \pi D^3 / 6$ for a particle of diameter $D$. The derivative of Eq. (4) is given by

$$\frac{dM(\xi)}{d\xi} = M_0 (\frac{1}{\xi^2} - \frac{1}{\sinh^2 \xi}). \quad (6)$$

In this study, we assumed that the external magnetic field at position **r** and time $t$ [$H(\mathbf{r}, t)$] is given by

$$H(\mathbf{r}, t) = H_S(\mathbf{r}) + H_D(t), \quad (7)$$

where $H_S(\mathbf{r})$ is the strength of the selection magnetic field at position **r** and $H_D(t)$ is the strength of the drive magnetic field at time $t$. In this study, we also assumed

$$H_D(t) = H_0 \cos(2\pi f_0 t), \quad (8)$$

where $H_0$ and $f_0$ denote the amplitude and frequency of the drive magnetic field, respectively. Thus, we obtain from Eqs. (3)-(8)

$$s(\mathbf{r},t) = -\frac{dM(\xi)}{d\xi} \cdot \frac{d\xi}{dt} = 2\pi f_0 M_0 H_0 \left(\frac{1}{\xi^2} - \frac{1}{\sinh^2 \xi}\right) \cdot \sin(2\pi f_0 t). \quad (9)$$

The total MPI signal in the field of view (FOV) at time $t$ [$u(t)$] is given by [9]

$$u(t) = \int_{FOV} s(\mathbf{r},t) c(\mathbf{r}) d\mathbf{r}, \quad (10)$$

where $c(\mathbf{r})$ denotes the concentration of MNPs at position $\mathbf{r}$. With the following Fourier decomposition:

$$s(\mathbf{r},t) = \sum_{n=0}^{\infty} S_n(\mathbf{r}) \cdot e^{i2\pi n f_0 t} \quad (11)$$

and

$$u(t) = \sum_{n=0}^{\infty} U_n \cdot e^{i2\pi n f_0 t}, \quad (12)$$

where $S_n(\mathbf{r})$ and $U_n$ denote the n-th frequency component of $s(\mathbf{r},t)$ and $u(t)$, respectively, and $i = \sqrt{-1}$, we obtain from Eq. (10)

$$U_n = \int_{FOV} S_n(\mathbf{r}) c(\mathbf{r}) d\mathbf{r}. \quad (13)$$

This means that every frequency component of the recorded MPI signal can be expressed as a weighted integral over the concentration in the imaging plane [11].

*2.3. Generation of field-free line*

According to Knopp et al. [12], an FFL can be generated by three or more Maxwell coil pairs. By varying the applied currents, the FFL can be arbitrarily rotated, while keeping the coils static in space. With additional Helmholtz coil pairs, the FFL can be translated [12].

The superposition of $L \geq 3$ (*L*: number of Maxwell coil pairs) ideal gradient fields rotated by equidistant angles ($\varphi_l = \pi \frac{l}{L}$) generates an FFL in the xy-plane through the center along direction $\mathbf{d}_{FFL}^{\alpha} = [\cos\alpha, \sin\alpha, 0]^T$ ($\alpha$: angle of FFL) for currents [12]

$$I_l = A\left[\frac{3}{2} - \cos(2\varphi_l - 2\alpha)\right], \qquad (14)$$

where *A* is given by

$$A = \frac{3}{2}\frac{G}{SL}. \qquad (15)$$

*G* and *S* in Eq. (15) denote the gradient strength in perpendicular direction to the FFL and a factor determined by the coil geometry, respectively [12]. Thus, FFL rotation can be easily achieved by varying the angle α.

For translating the FFL to spatial position **r**, the total magnetic field at this position [$H(\mathbf{r})$] has to be canceled out, that is,

$$H(\mathbf{r}) = H^{\alpha}(\mathbf{r}) + H_{trans} = 0, \qquad (16)$$

where $H^\alpha(\mathbf{r})$ denotes the magnetic field at position **r** when the FFL was rotated by α, and $H_{trans}$ denotes the magnetic field for translation. From Eq. (16), we obtain [12]

$$H_{trans} = -H^\alpha(\mathbf{r}).  \qquad (17)$$

*2.4. Generation of projection data*

The projection data at position $r$ and angle $\theta$ [$P(r, \theta)$] was generated by

$$P(r,\theta) = \sum_{n=1}^{\infty} U_n(r,\theta), \qquad (18)$$

where $U_n(r,\theta)$ denotes the n-th frequency component of the MPI signal given by Eq. (12) when the distance to the FFL from the origin is $r$ and the angle of FFL is $\theta$, as illustrated in Fig. 1.

*2.5. Image reconstruction*

Image reconstruction was performed using the FBP method [4] and the ML-EM algorithm [5].

With the FBP method, the concentration of MNPs at $(x, y)$ [$c(x, y)$] is calculated by

$$c(x,y) = \int_0^\pi \tilde{P}(-x\cos\theta + y\sin\theta, \theta)d\theta, \qquad (19)$$

where $\tilde{P}(r,\theta) = P(r,\theta) \otimes g(r)$, with $g(r)$ being a reconstruction filter and $\otimes$ being convolution integral.

With the ML-EM algorithm, the concentration of MNPs at position $j$ and after $k+1$ iterations ($c_j^{k+1}$) is given by [5]

$$c_j^{k+1} = \frac{c_j^k}{\sum_i p_{ij}} \sum_i \frac{p_{ij} P_i}{\sum_k p_{ik} c_k^n}, \tag{20}$$

where $i$ denotes a projection bin (Fig. 1), $P_i$ is the projection data at bin $i$ (Fig. 1), $p_{ij}$ is the probability that an MPI signal at pixel $j$ contributes to projection bin $i$. It should be noted that $p_{ij}$ was assumed to be unity when both the projection bin $i$ and pixel $j$ are located on the same FFL as illustrated in Fig. 1, otherwise $p_{ij}$ was assumed to be zero in this study.

*2.6. Simulation studies*

Figure 2 illustrates the phantom used in simulation studies. The phantom was discretized on a rectangular grid of matrix size $128 \times 128$ ($128 \times 128$ mm$^2$) and consists of 18 holes with different diameters ranging from 1 mm to 16 mm, filled with MNPs.

In this study, we considered magnetite (Fe$_3$O$_4$) as MNPs and $M_d$ in Eq. (5) was taken as 446 kA/m [13]. Simulation studies were performed under the following conditions. When using the FBP method, a Shepp-Logan filter [14] was used. When using the ML-EM algorithm, the number of iterations was taken as 32. The number of Maxwell coil pairs (*L*), the frequency of drive magnetic field ($f_0$), and SNR were fixed at 4, 25 kHz, and 10, respectively, in all studies. Unless specifically stated, the number of projection data, the gradient of selection magnetic field, the strength of drive magnetic field, and the diameter of MNPs were taken as 32, 5 T/m, 5 mT, and 30 nm, respectively. When investigating the effect of the gradient of selection magnetic field, the gradient strength was varied from 2 T/m to 10 T/m with an increment of 2 T/m (Fig. 4). When investigating the effect of the diameter of MNPs, it was varied from 10 nm to 50 nm with an increment of 10 nm (Fig. 5). When investigating the effect of the number of projection data, it was varied as 4, 8, 16, 32, and 64 (Fig. 6). When investigating the effect of the strength of drive magnetic field, it was varied as 1, 5, 10, 20, and 30 mT (Fig. 7).

## 3. Results

Figure 3(a) shows an example of the FFL rotated by 0, 30, 60, 90, and 120 degrees, while Fig. 3(b) shows cases when the FFL was rotated by 60 degrees and translated by 0, 20, 40, -20, and -40 pixels.

Figure 4 shows the comparison of the images reconstructed by the FBP method and the ML-EM algorithm for various gradient of selection magnetic field. The number in the figure represents the gradient of selection magnetic field in mT. As shown in Fig. 4, the spatial resolution of the reconstructed images became better with increasing gradient of selection magnetic field. In general, the image quality was improved by using the ML-EM algorithm compared to the FBP method. Especially, streak artifacts were reduced by use of the ML-EM algorithm.

Figure 5 shows the comparison of the images reconstructed by the FBP method and the ML-EM algorithm for various diameters of MNPs. The number in the figure represents the diameter of MNPs in nm. As shown in Fig. 5, the spatial resolution of the reconstructed images became better with increasing diameter of MNPs up to approximately 30 nm. For the diameter greater than approximately 30 nm, the spatial resolution was almost constant from visual inspection.

Figure 6 shows the comparison of the images reconstructed by the FBP method and the ML-EM algorithm for various numbers of projection data. The number in the figure represents the number of projection data. As shown in Fig. 6, the image quality of the reconstructed images became better with increasing number of projection data. The image quality was considerably improved by using the ML-EM algorithm compared to the FBP method especially when the number of projection data was small.

Figure 7 shows the comparison of the images reconstructed by the FBP method and the ML-EM algorithm for various strength of drive magnetic field. The number in the figure represents the strength of drive magnetic field in mT. As shown in Fig. 7, the spatial resolution of the reconstructed images became worse with increasing strength of drive magnetic field. When the strength of drive magnetic field decreased, there was a tendency for the reconstructed images to become noisy especially when using the FBP method. On the other hand, when using the ML-EM algorithm, the streak artifacts observed in the images reconstructed using the FBP method were reduced, even when the strength of drive magnetic field was weak.

## 4. Discussion

In this study, we presented simulation studies on the image reconstruction in MPI with an FFL encoding scheme. Since the FFL encoding scheme is similar to the image encoding scheme used in x-ray CT, we expected that the image reconstruction method used in x-ray CT is applicable to MPI with this encoding scheme. Then, we investigated the feasibility of the FBP method in MPI. Furthermore, we also proposed the use of the ML-EM algorithm for improving the image quality of MPI. To the best of our knowledge, there have been no reports on these studies, and the use of the ML-EM algorithm in MPI has not previously been reported in the literature. Our preliminary results (Figs. 4-7) demonstrated the feasibility of imaging the spatial distribution of MNPs using MPI with an FFL encoding scheme and the image reconstruction methods used in x-ray CT or nuclear medicine. Our results also suggested that the ML-EM algorithm is useful for improving the image quality, especially when the strength of drive magnetic field is weak (Fig. 7) and the number of projection data is small (Fig. 6). These results suggest that with the use of the ML-EM algorithm, data acquisition time in MPI can be reduced by decreasing the number of projection data.

The ML-EM algorithm has several attractive features [5]: the reconstructed images are non-negative; the total image density is preserved at each iteration; the convergence is assured in theory; and the reconstructed images converge to the maximum likelihood estimate. However, the application of this algorithm is computer intensive and convergence slow. To overcome these drawbacks, Hudson and Larkin [6] developed the OS-EM algorithm as an accelerated version of the ML-EM algorithm. The OS-EM algorithm processes the projection data in subsets within each iteration and this procedure accelerates convergence compared to the ML-EM algorithm by a factor proportional to the number of subsets [6]. The ML-EM algorithm can be considered as a particular case when a single subset includes all projection data. Thus, the OS-EM algorithm will be used instead of the ML-EM algorithm for practical use.

In this study, $p_{ij}$ in Eq. (20) was assumed to be unity when both the projection bin $i$ and pixel $j$ are located on the same FFL, otherwise $p_{ij}$ was assumed to be zero. If we can include the effect of the position-dependent sensitivity of the receive coil in $p_{ij}$ in Eq. (20), the accuracy of the reconstructed images would be further improved.

As previously described, an FFL can be generated by three or more Maxwell coil pairs and can be translated by additional Helmholtz coil pairs [12].

The present simulation studies were based on the assumption of idealized magnetic gradient fields. That is, we assumed that the Maxwell coil pairs have infinite distance and diameter. For finite distance and diameter, the approximation is only valid in a certain region at the center between both coils. Therefore, as pointed out by Knopp et al. [12], the magnetic field of each Maxwell coil pair is linear only for the high symmetric point and deviates in radial direction from the ideal case. In order to examine the accuracy of the FFL for real magnetic fields, further studies should be carried out using numerical evaluation of the Biot-Savart law [15].

The spatial resolution ($\Delta x$) of one-dimensional MPI system has been derived by Rahmer et al. [9] as

$$\Delta x \approx \frac{4 k_B T}{\mu_0 M_d V_M G} \approx \frac{24 k_B T}{\pi \mu_0 M_d D^3 G}, \qquad (21)$$

where $D$ and $G$ are the diameter of MNPs and the gradient strength of selection magnetic field, respectively. Thus, $\Delta x$ is inversely proportional to $D^3$ and $G$. As shown in Fig. 4, the spatial resolution of the reconstructed images became better with increasing $G$. Furthermore, the spatial resolution of the reconstructed images became better with increasing $D$ up to approximately 30 nm (Fig. 5). These results agree with those expected from Eq. (21). However, when $D$ was greater than approximately 30 nm, the spatial resolution was almost constant from visual inspection (Fig. 5), which is different from expectation. Although the reason for this finding is not clear, this may be due to the fact that Eq. (21) was derived from the analysis of one-dimensional MPI system, and Eq. (21) may not always be applicable to two-dimensional MPI system. Furthermore, we found that the spatial resolution of MPI largely depends on the strength of drive magnetic field and there may be an optimum value for this parameter (Fig. 7). Our simulation studies may also be useful for optimizing these parameters when designing an MPI system.

In conclusion, we presented image reconstruction methods for MPI with an FFL encoding scheme. Our preliminary results demonstrated that the FBP method or the ML-EM algorithm is applicable to image reconstruction in MPI with an FFL encoding scheme, and suggested that the ML-EM algorithm is useful for improving the image quality of MPI.

**Figure Legends**

Figure 1. Illustration of a coordinate system to mathematically describe the image reconstruction problem in magnetic particle imaging with a field-free line (FFL) encoding scheme. $P(r,\theta)$ represents the projection data when the distance to the FFL from the origin is $r$ and the angle of FFL is $\theta$. $P_i$ represents the projection data at projection bin $i$.

Figure 2. Phantom used in simulation studies. The phantom is sized $128 \times 128$ mm$^2$ containing $128 \times 128$ pixel. The number in the figure represents the diameter of the hole in mm, which is assumed to be filled with magnetic nanoparticles (MNPs).

Figure 3. (a) An example of the FFL rotated by 0, 30, 60, 90, and 120 degrees. (b) An example of the FFL rotated by 60 degrees and translated by 0, 20, 40, -20, and -40 pixels.

Figure 4. Comparison of the images reconstructed by the filtered backprojection (FBP) method with a Shepp-Logan filter and the maximum likelihood-expectation maximization (ML-EM) algorithm with an iteration number of 32 for various gradient of selection magnetic field. The number in the figure represents the gradient of selection magnetic field in mT. In these cases, the signal-to-noise ratio (SNR), the number of projection data, the strength of drive magnetic field, and the diameter of MNPs were taken as 10, 32, 5 mT, and 30 nm, respectively.

Figure 5. Comparison of the images reconstructed by the FBP method and the ML-EM algorithm for various diameters of MNPs. The number in the figure represents the diameter of MNPs in nm. In these cases, SNR, the number of projection data, the strength of drive magnetic field, and the gradient of selection magnetic field were taken as 10, 32, 5 mT, and 5 T/m, respectively.

Figure 6. Comparison of the images reconstructed by the FBP method and the ML-EM algorithm for various numbers of projection data. The number in the figure represents the number of projection data. In these cases, SNR, the strength of drive magnetic field, the gradient of selection magnetic field, and the diameter of MNPs were taken as 10, 5 mT, 5 T/m, and 30 nm, respectively.

Figure 7. Comparison of the images reconstructed by the FBP method and the ML-EM algorithm for various strength of drive magnetic field. The number in the figure represents the strength of drive magnetic field in mT. In these cases, SNR, the number of projection data, the gradient of selection magnetic field, and the diameter of MNPs were taken as 10, 32, 5 T/m, and 30 nm, respectively.

**Figures**

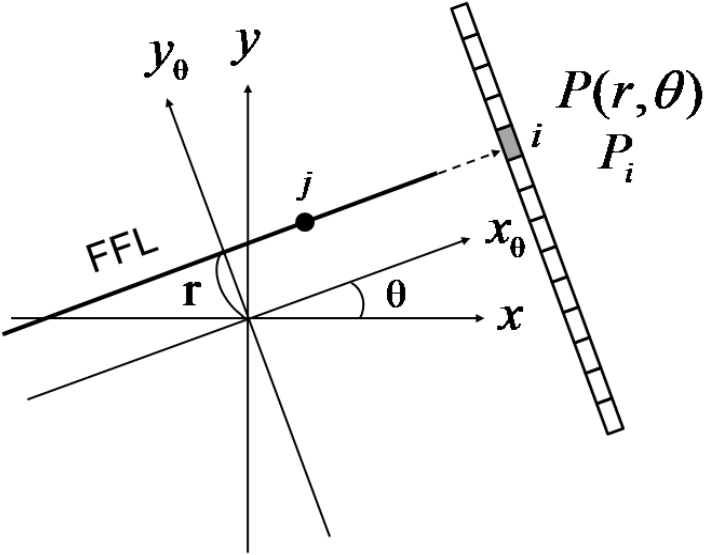

**Figure 1**

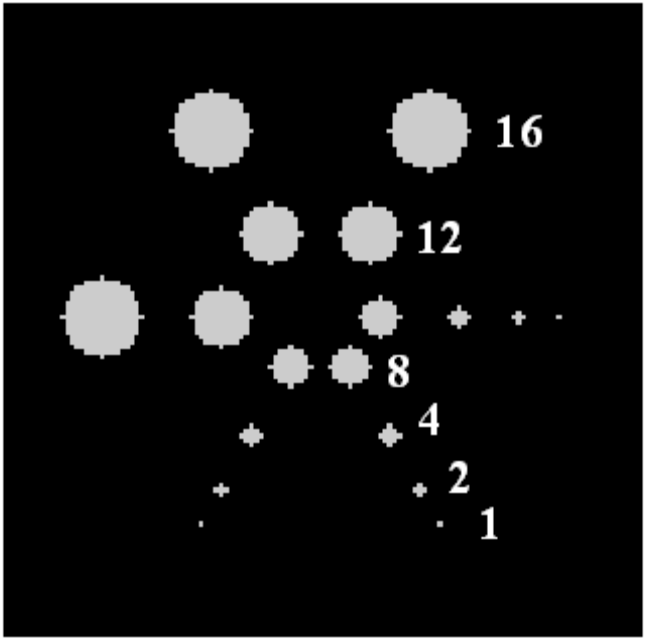

**Figure 2**

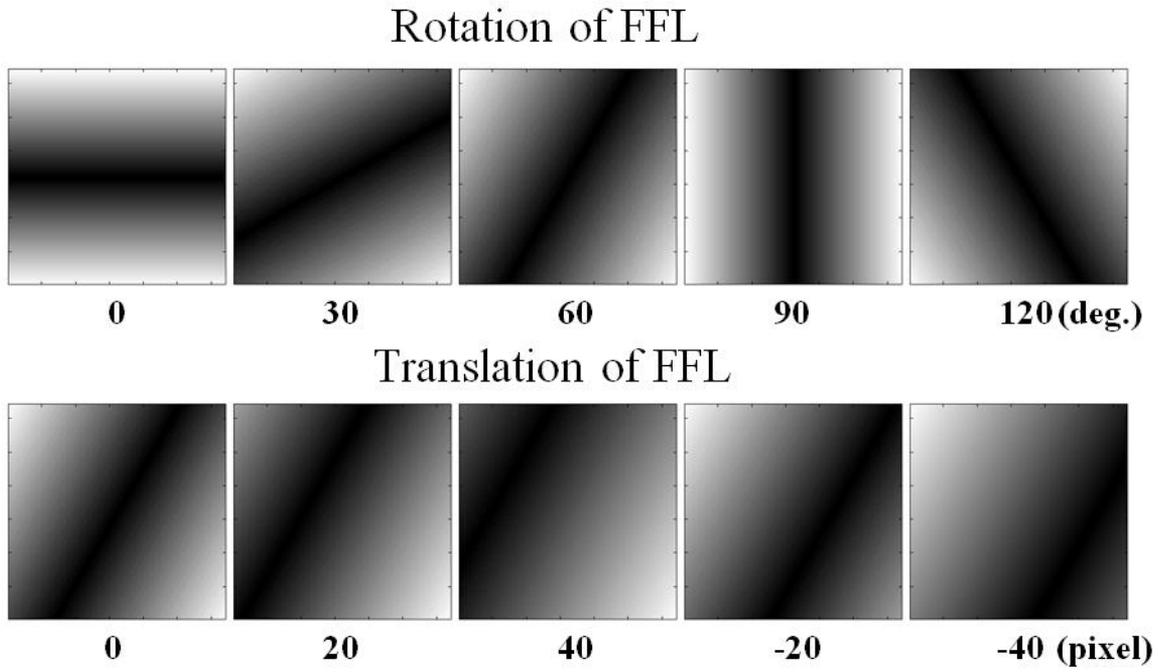

**Figure 3**

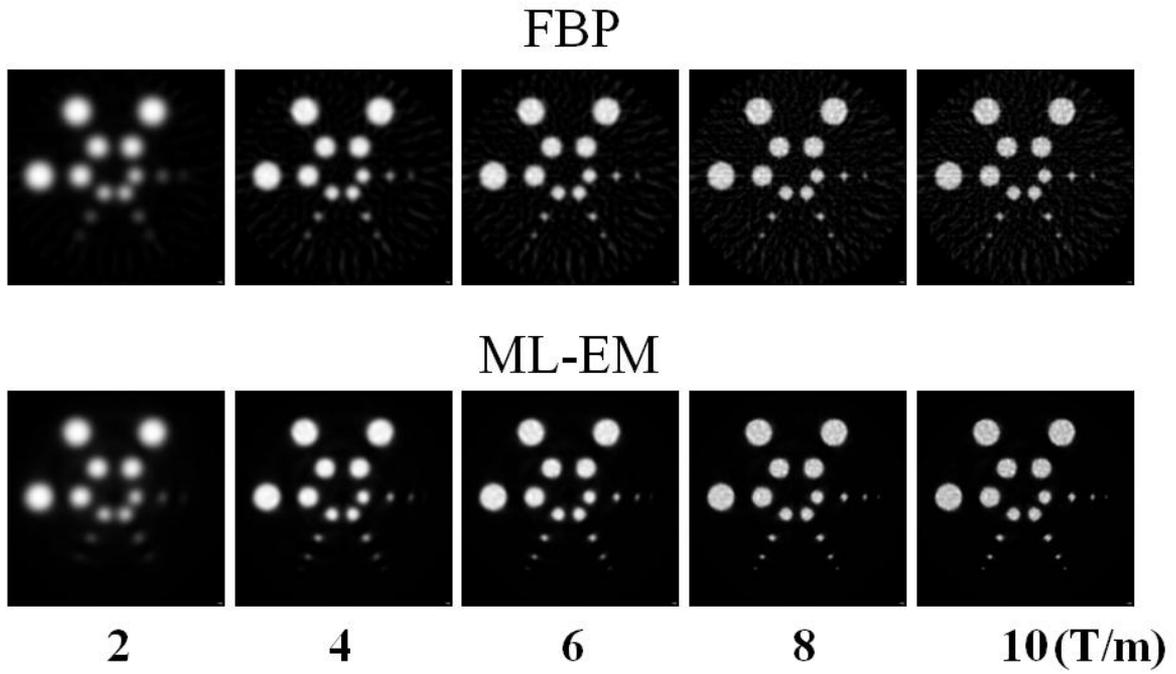

**Figure 4**

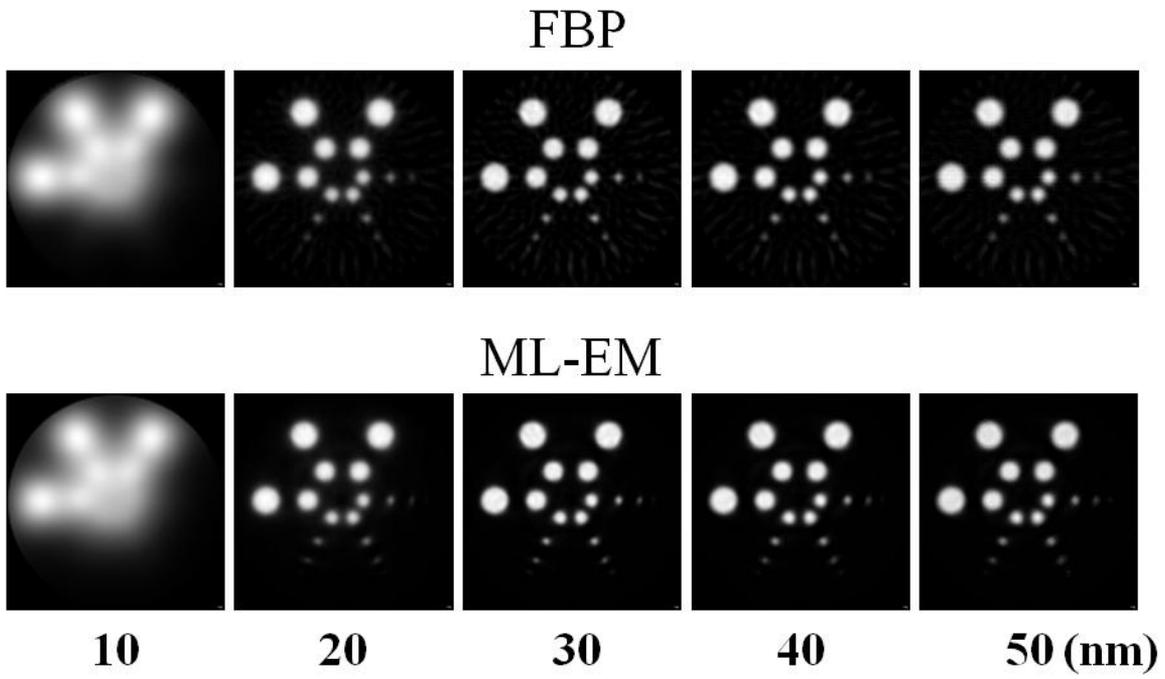

Figure 5

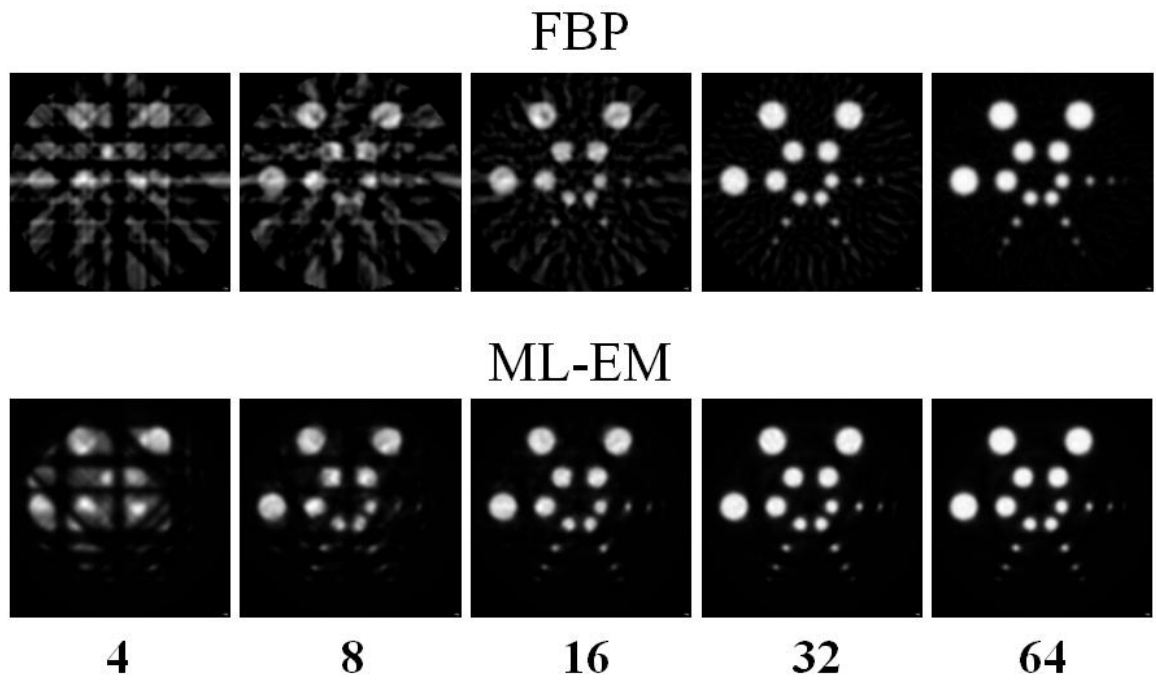

Figure 6

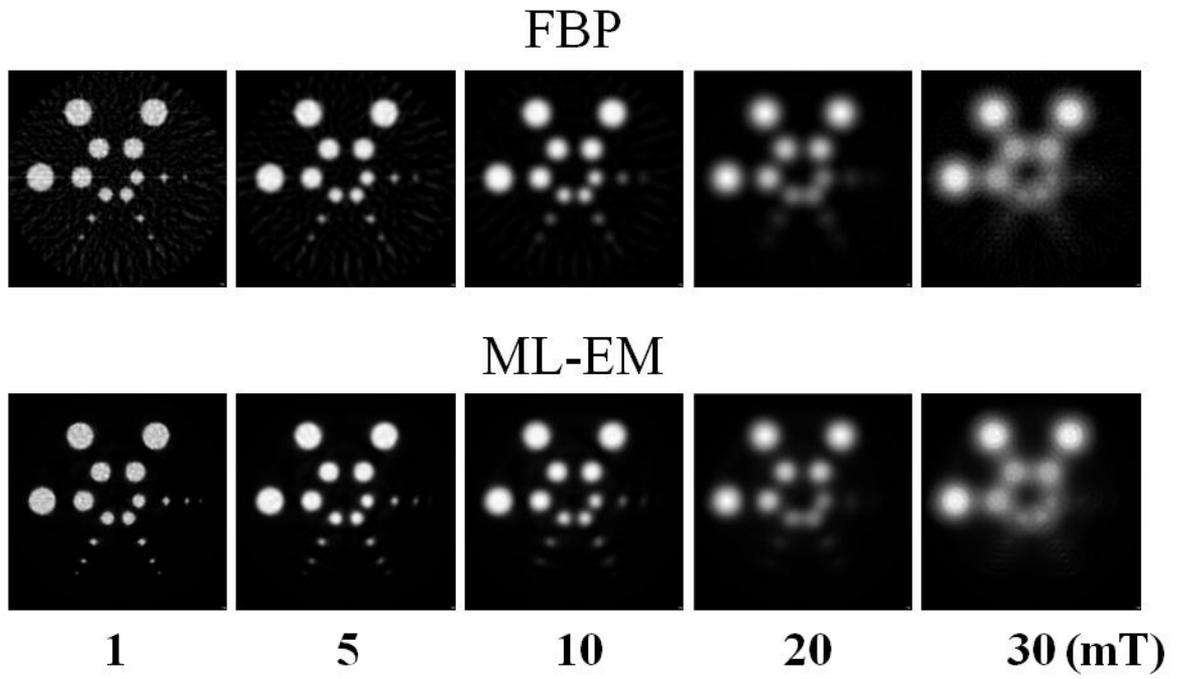

**Figure 7**